\documentclass[a4paper,11pt]{article}
\pdfoutput=1 

\usepackage{jinstpub} 

\title{\boldmath Monte-Carlo study of the MRPC prototype for the upgrade of BESIII}

\author[a,1]{F.F. An, \note{Corresponding author.}}
\author[a]{S.S. Sun, }
\author[a]{H.M. Liu, }
\author[a]{W.G. Li, }
\author[a]{Z.Y. Deng, }
\author[a]{H.H. Liu, }
\author[a]{J.Y. Liu }
\author[b]{and R.X. Yang}

\affiliation[a]{Institute of High Energy Physics, Chinese Academy of Science,
  \\Beijing 100049, China}
\affiliation[b]{University of Science and Technology of China, Hefei, \\Anhui 230026, China}

\emailAdd{anff@ihep.ac.cn}

\abstract{
A GEANT4-based simulation  is developed  for the endcap 
time of flight (ETOF) upgrade 
based on multi-gap resistive plate chambers (MRPC)
for the BESIII experiment.
The MRPC prototype and the simulation method are described.
Using a full Monte-Carlo simulation, the influence of high voltage and threshold on 
time resolution and detection efficiency are investigated. 
The preliminary results from simulation are presented
and are compared with the experimental data
taken with the prototype MRPC modules.
}


\proceeding{13$^{\text{th}}$ Workshop on Resistive Plate Chambers and Related Detectors (RPC2016)\\
  Feb. 22-26, 2016\\
  Ghent, Belgium}

\begin{document}
\maketitle
\flushbottom

\section{Introduction}
\label{sec:intro}

The BESIII experiment~\cite{besiii} is designed to study 
 $\tau$-charm physics ($\sqrt{s}$=2-4.6 GeV) in $e^{+}e^{-}$ collisions at the 
double-ring Beijing Electron-Positron Collider (BEPCII).
The time-of-flight (TOF) subdetector of BESIII plays an important
role in particle identification (PID).
In 2013, BESIII started an endcap TOF (ETOF) upgrade based on 
multi-gap resistive plate chambers (MRPC), which could effectively 
reduce the influence of multiple scattering effects 
with the main drift chamber (MDC) 
\cite{Hui:2013yea},
and significantly improve the PID ability~\cite{Ullrich:2014gba}.
In 2015, two prototype MRPC modules were installed  
for performance  check, and participated in  data taking during physics runs
under a series of high voltages and thresholds.

Monte-Carlo (MC) simulation is necessary in understanding the behavior of
the MRPC detector and minimizing systematic uncertainties 
in the future physics analysis at the BESIII experiment.
Therefore a reliable simulation software is desirable
to describe the new detector and physical interactions.
A simulation package has been built based on GEANT4~\cite{geant}.
It has been implemented into the BESIII 
Offline Software System (BOSS)~\cite{Li:2006yea},
capable to perform full simulation and reconstruction.
 In this paper, the MRPC prototype and the simulation method are described in detail.
 The performance is studied related with time resolution and detection efficiency based 
on the experimental data taken with the prototype modules.

\section{Experiment setup}
\label{sec:geom}

\begin{figure}[htp]
\centering
\subfigure[]
{
\label{fig:endcap}
\includegraphics[width=0.25\linewidth]{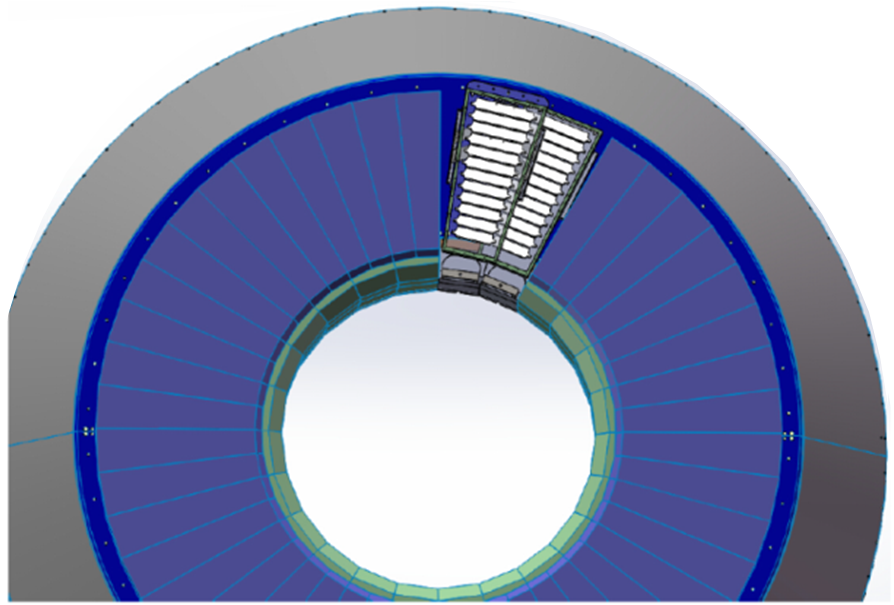}
\hspace{0.015\linewidth}
}
\subfigure[]
{
\label{fig:bare_chamber}
\includegraphics[width=0.39\linewidth]{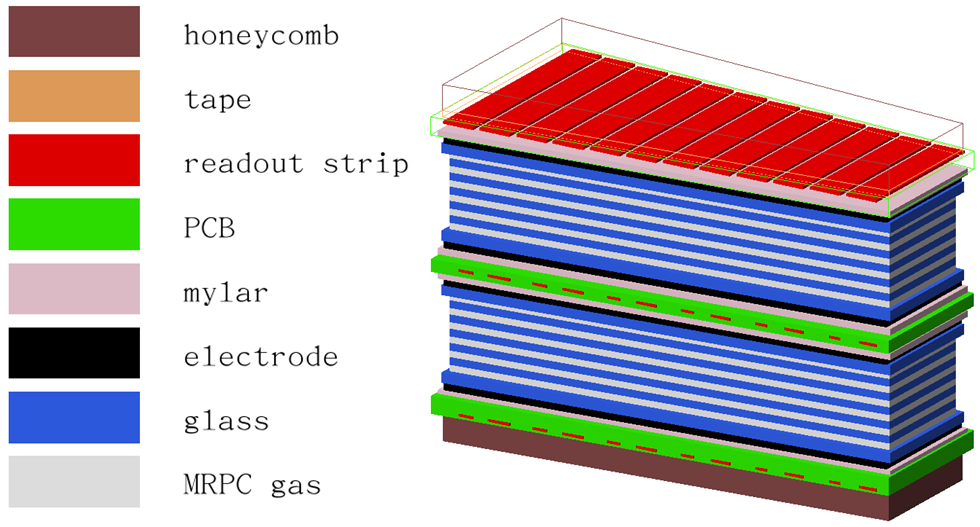}
\hspace{0.01\linewidth}
}
\subfigure[]
{
\label{fig:module}
\includegraphics[width=0.2\linewidth]{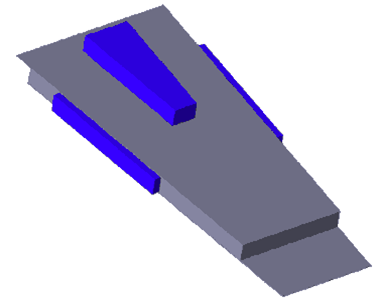}
}
\caption{\label{fig:geom}
Experimental setup for MRPC performance check:
(a) One TOF endcap with two MRPC modules installed; 
(b) Schematic representation of one MRPC bare chamber;
(c) A single MRPC module.}
\end{figure}

The BESIII detector is a general-purpose spectrometer consisting of five main components 
arranged cylindrically symmetric around the interaction point. 
Starting from the innermost, it consists of a MDC,
a scintillator TOF, a crystal electromagnetic calorimeter (EMC),
a superconducting solenoid magnet surrounding the EMC barrel
with an axial uniform magnetic field of 1.0 T, and an
outermost  muon chamber (MUC) embedded
within the return yoke of the magnet. The ETOF system is mounted inside 
the EMC endcap. Two prototype MRPC modules are installed into ETOF,
taking place of four scintillators with a coverage of 1/18 in azimuth,
as shown in Figure~\ref{fig:endcap}.

The design of one MRPC module is shown in Figure~\ref{fig:bare_chamber}.
It has 12 gas gaps arranged in a double stack. The gaps are 0.22 mm thick 
separated by 0.4 mm thick inner glasses, which are used as resistive plates 
with volume resistivity of about $10^{13}$ $\Omega$ cm.
The  0.55 mm thick outer  glasses 
are coated with graphite film, which is used as the electrodes.
The electrodes are separated from the readout strips by mylar films.
 In each module, there are 12 double-ended readout strips embedded in 
 printed circuit boards (PCB),
each of which is 25 mm wide with a 4 mm interval between them.
PCBs are covered with honeycomb boards to hold the whole structure.
The module is enclosed in a 25 mm thick gas-tight aluminum box shown 
 in Figure~\ref{fig:module}, which is filled with a gas mixture of
 90\% C$_{2}$F$_{4}$H$_{2}$, 5\% SF$_{6}$ and 5\% iso-C$_{4}$H$_{10}$.  
The front-end electronics (FEE) are contained in the additional boxes at both sides 
and on the top of the aluminum box.

\section{Simulation method}
\label{sec:sim}

The exact description of the geometry and material and the physics 
processes involving the interaction between the incident particles and 
the working gas, are realized using GEANT4.
The main  processes in simulation consist of primary ionization, electron multiplication, 
signal induction, and conversion from charge signals to time information.
These processes have been 
discussed in Ref.~\cite{Riegler:2003yea, Shao:2006yea, Ullrich:2014gba}. Here we mainly 
present some primary results related to each process.

\subsection{Primary ionization}
\label{sec:ion}

\begin{figure}[htp]
\centering
\includegraphics[width=0.4\linewidth]{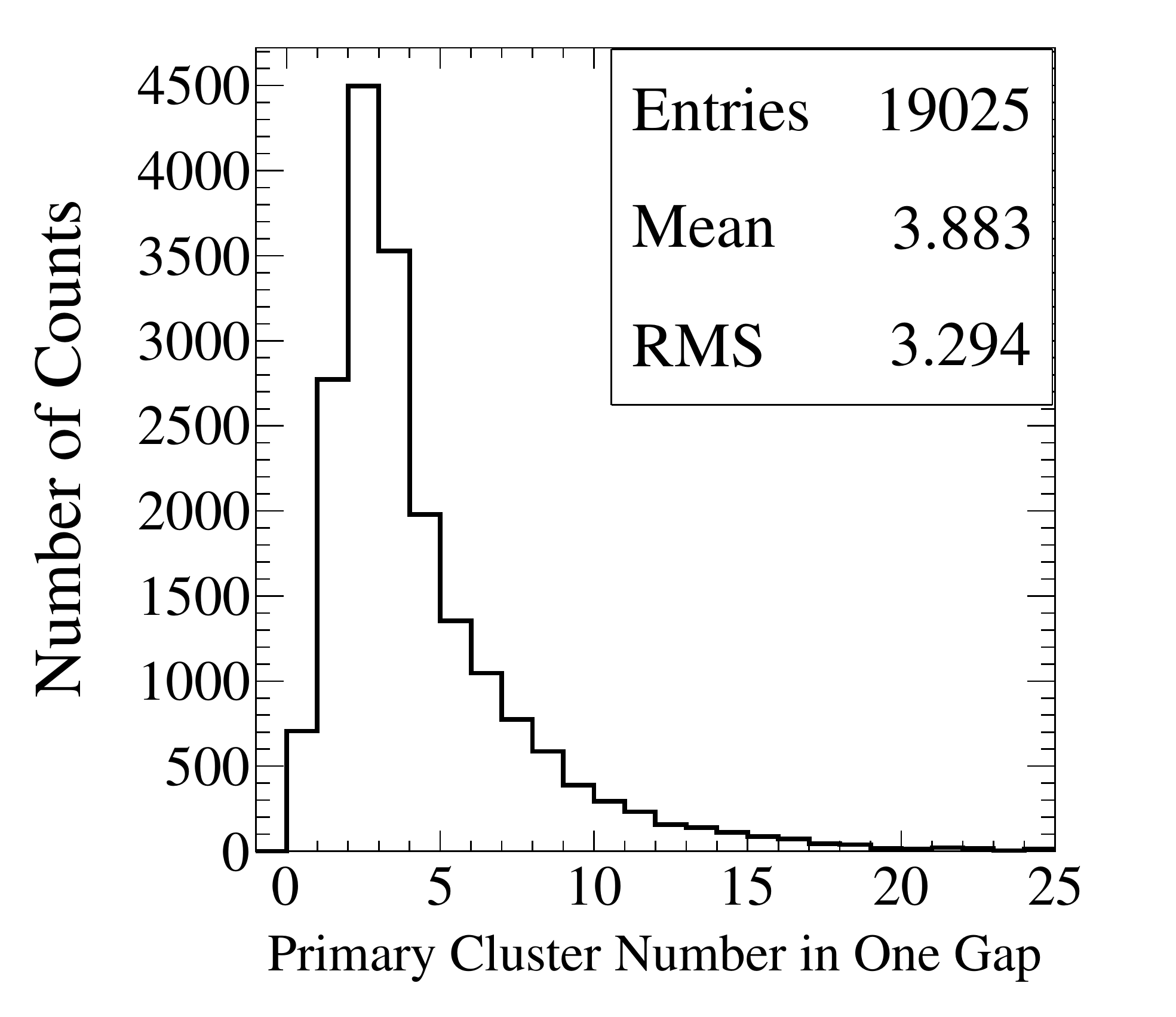}
\qquad
\includegraphics[width=0.4\linewidth]{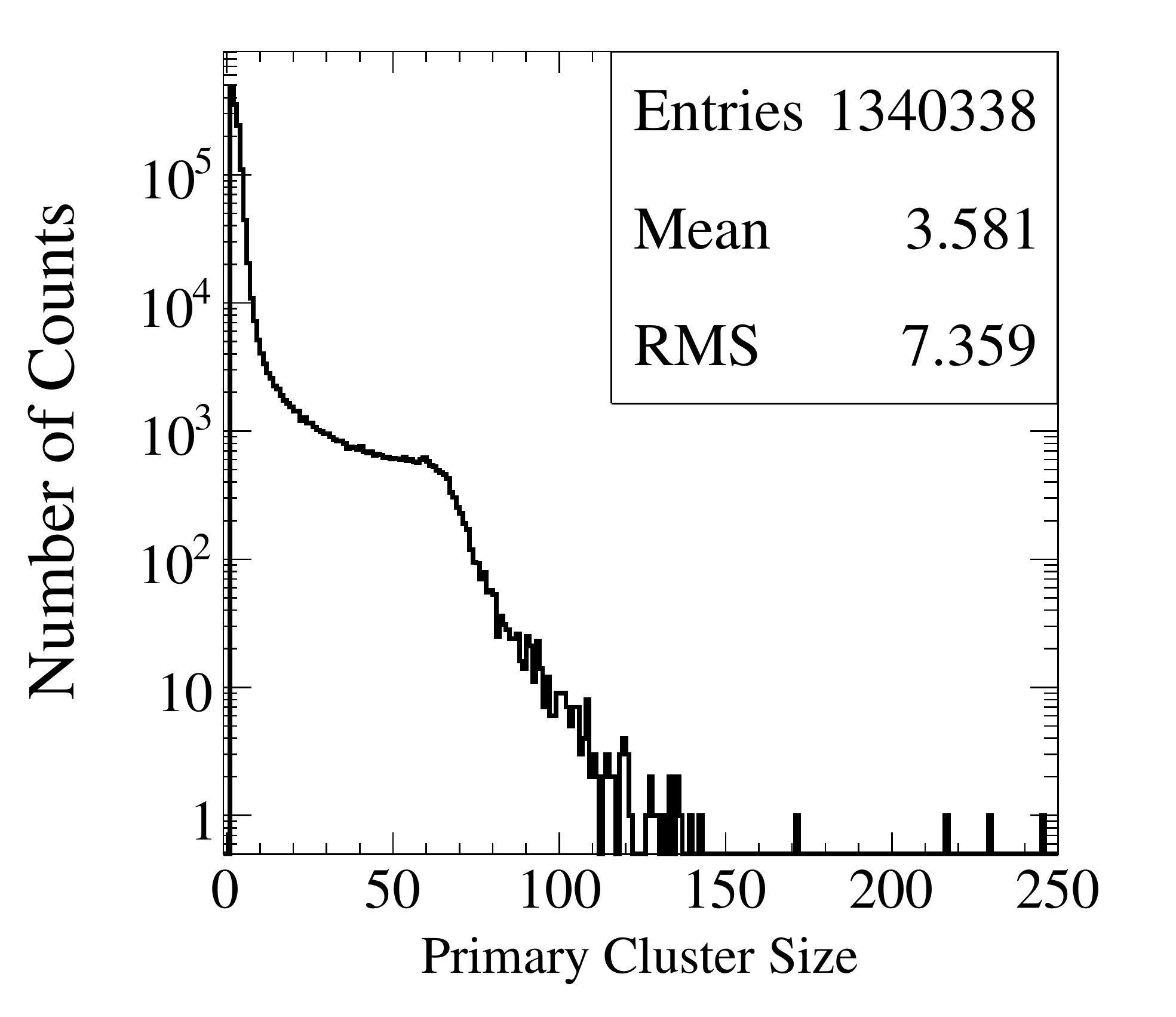}
\caption{\label{fig:ion} 
Left: the number of primary clusters in one gap.
Right: the number of primary electrons in one cluster.}
\end{figure}

The primary ionization is characterized by the average number of clusters 
per unit length and the probability distribution of the number of 
electrons per cluster. 
In simulation these parameters of the working gas are determined by GEANT4.
The simulation results are  illustrated in Figure~\ref{fig:ion}:
 there is an average of about 4 clusters in 
one gap, and each cluster most likely starts from 3-4 primary electrons.

\subsection{Electron multiplication}
\label{sec:multi}

\begin{figure}[htp]
\centering
\subfigure[]
{
\label{fig:gas_coef}
\includegraphics[width=0.38\linewidth]{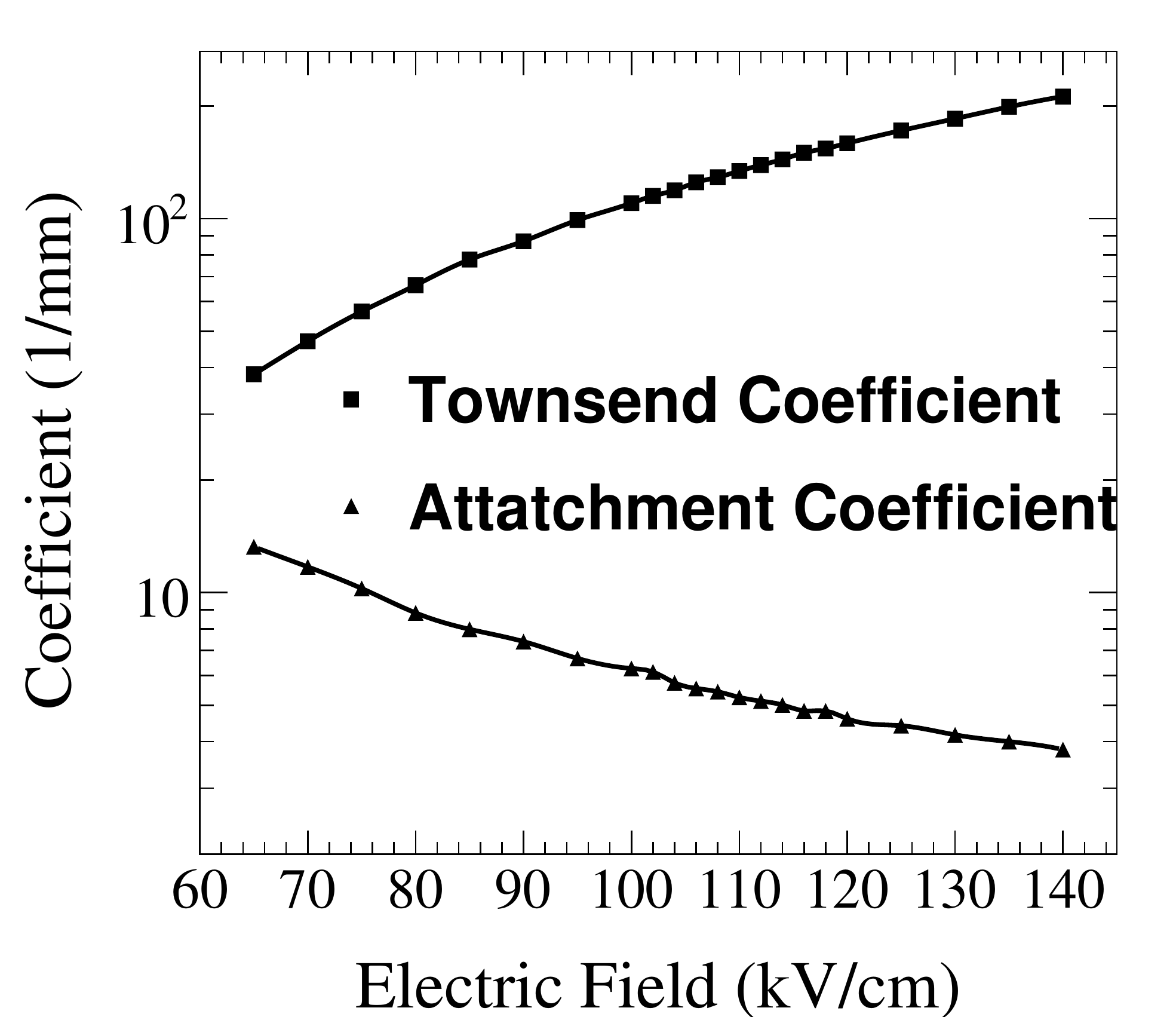}
\qquad
}
\subfigure[]
{
\label{fig:avalanche}
\includegraphics[width=0.38\linewidth]{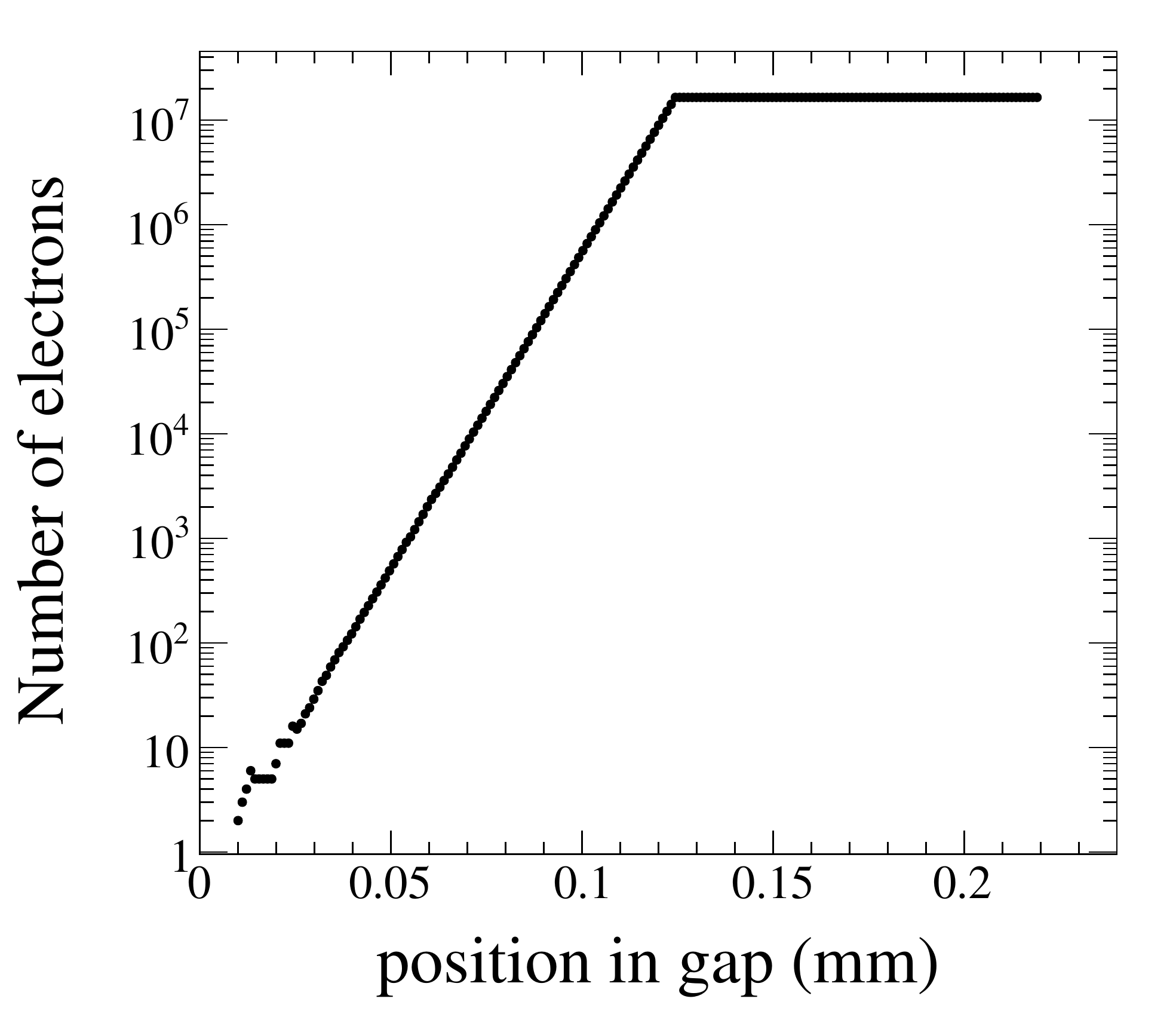}
}
\caption{
\label{fig:multi}
(a) Townsend and attachment coefficient
 calculated by MAGBOLTZ for the 
working gas (90\% C$_{2}$F$_{4}$H$_{2}$, 5\%
 SF$_{6}$ and 5\% iso-C$_{4}$H$_{10}$) with the temperature
 and pressure at $25\,^{\circ}\mathrm{C}$ and 1 atm, respectively;
(b) development of an avalanche starting from 0.01 mm.}
\end{figure}

Two parameters are used to characterize the multiplication process:
one is  the Townsend coefficient $\alpha$ describing the 
probability of multiplication due to secondary ionization,
and the other is the attachment coefficient $\eta$ describing the 
probability of attachment. The variation of both coefficients with 
electric field  are calculated using MAGBOLTZ~\cite{Biagi:1999yea},
as is shown in Figure~\ref{fig:gas_coef}
for the working gas under standard temperature and pressure.

Figure~\ref{fig:avalanche} shows the development of one avalanche
along the gas gap starting from 0.01 mm.
It grows almost exponentially at the early stage, and stops growing
when the avalanche contains electrons more than 
$N_{sat}$ (1.5$\times 10^{7}$ in this study, close to the limit for avalanche-streamer
transition~\cite{Fonte:1991yea}) for  consideration about 
the space charge effect. 
An accurate simulation  of the space charge effect involving the dynamic 
calculation of the electric field contributed by the avalanche charges
is investigated in Ref.~\cite{Lippmann:2004yea}.
In this study, a simple cut-off is used because the time resolution 
and detection efficiency which really concern us 
are only sensitive to the early stage of an avalanche where the space charge 
effect can be neglected.  We vary the value of $N_{sat}$
and observe no obvious change in the resolution and efficiency of the  collision data.

\subsection{Signal induction}
\label{sec:signal}

\begin{figure}[htp]
\centering
\subfigure[]
{
\label{fig:gas_v}
\includegraphics[width=0.38\linewidth]{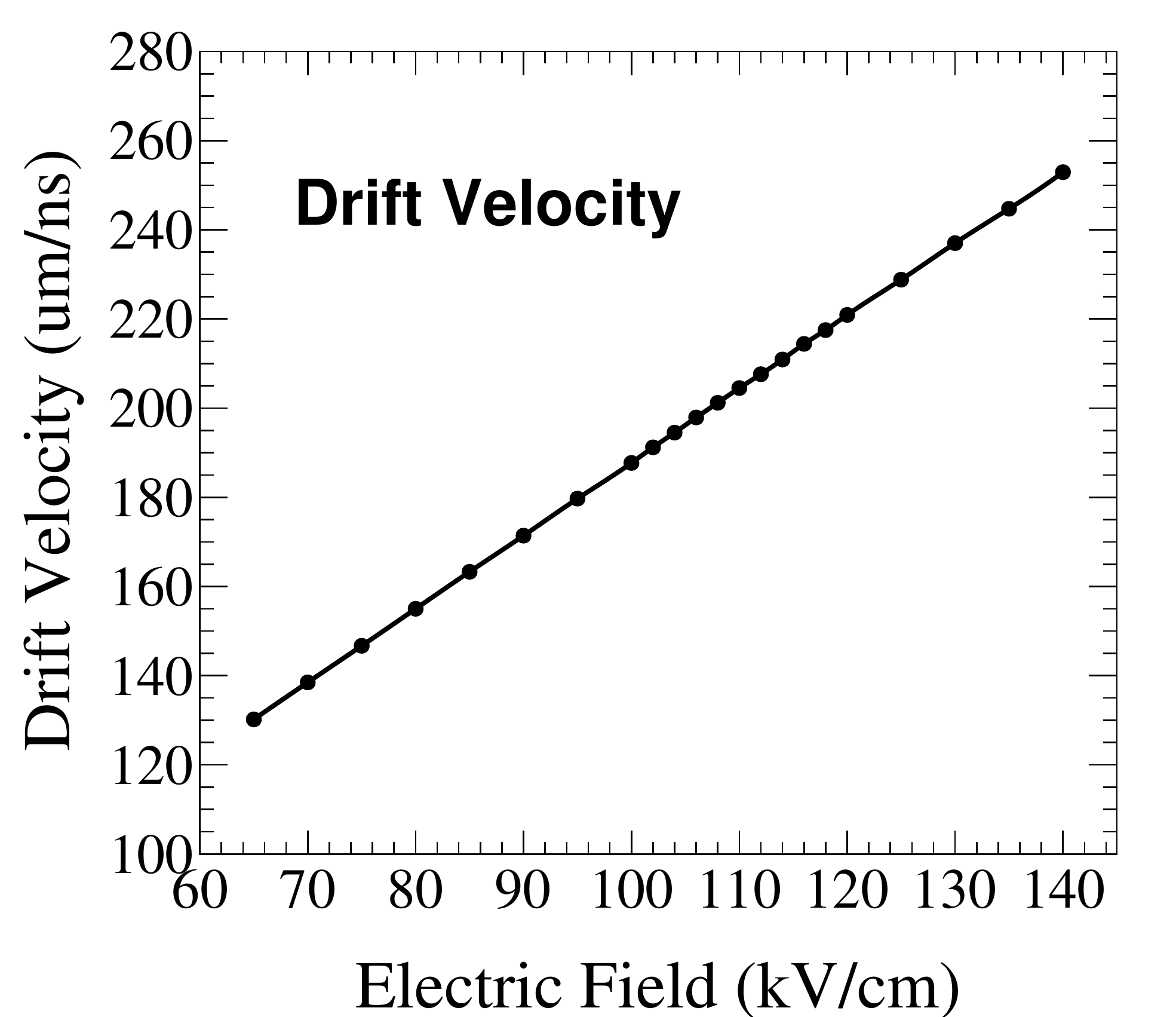}
\qquad
}
\subfigure[]
{
\label{fig:charge}
\includegraphics[width=0.38\linewidth]{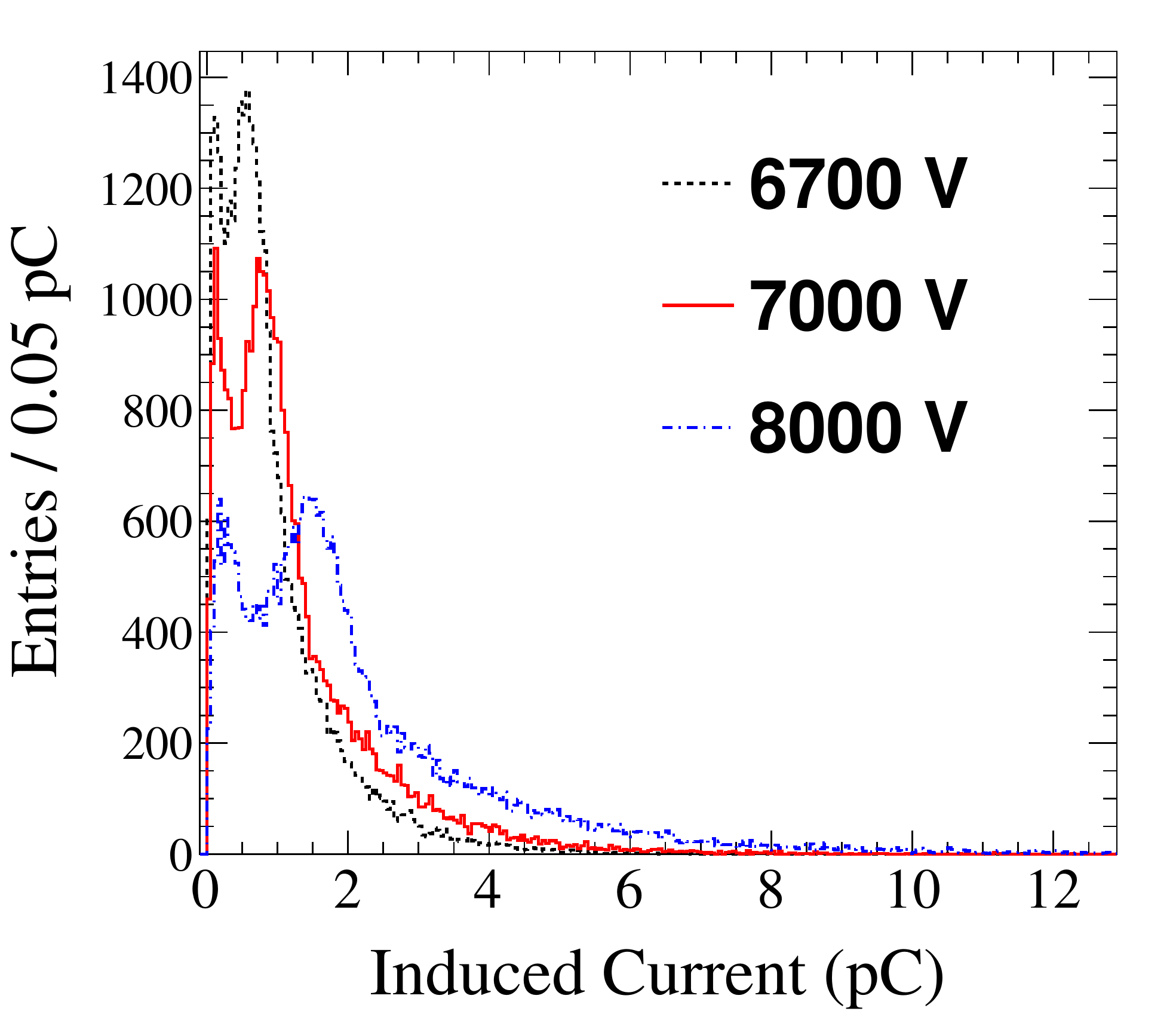}
}
\caption{\label{fig:signal}
(a) Drift velocity calculated by MAGBOLTZ for the 
working gas (90\% C$_{2}$F$_{4}$H$_{2}$, 5\%
 SF$_{6}$ and 5\% iso-C$_{4}$H$_{10}$) with the temperature
 and pressure at $25\,^{\circ}\mathrm{C}$ and 1 atm, respectively;
(b) induced charge spectra under different electric fields.}
\end{figure}

The signal caused by positive and negative 
ions is neglected because their velocities  are very small.
The induced current  from electrons is calculated
 using the Ramo's theorem~\cite{Ramo:1939yea}:

\noindent
\begin{equation}
\label{eq:signal}
i(t) = E_{\rm weight}v_{\rm drift}Q_{e}N(t),
\end{equation}

\noindent
with $E_{\rm weight}$ being the weighting field~\cite{Shao:2006yea}, 
$v_{\rm drift}$ the drift velocity, 
$Q_{e}$  the electron charge,
and $N(t)$  the number of electrons at time $t$.
The  $v_{\rm drift}$ variation with the electric field
 is calculated using MAGBOLTZ,  as is shown in Figure~\ref{fig:gas_v}.
The induced charge spectra under different electric fields are illustrated in 
Figure~\ref{fig:charge}. 
The peak of the charge spectrum moves towards 
the right side with the increase of the  electric field. Each 
spectrum has double peaks:
the right peak represents the most possible charge,
while the left one is caused by the
hits at the chamber  boarders or under the strip intervals.
Under the electric field of 7000 V, we have a spectrum 
peaking around 1 pC.

\subsection{Signal charge to time conversion}
\label{sec:q2t}

The signal information measured by the detector is the 
absolute timing of the leading  edge (TDC)
and the time over threshold (TOT).
TDC is considered as the sum of the 
flight time of the particle ($t_{\rm flight}$), the threshold crossing time of the avalanche  
($t_{\rm thres}$) and the propagation time along the strip ($t_{\rm prop}$).
TOT denotes the delay between the leading-edge and trailing-edge timing,
which is used in the time slewing correction to remove the 
timing uncertainty introduced by fixed-threshold leading-edge 
discrimination in the offline calibration.

\begin{figure}[htp]
\centering
\subfigure[]
{
\label{fig:thres}
\includegraphics[width=0.304\linewidth]{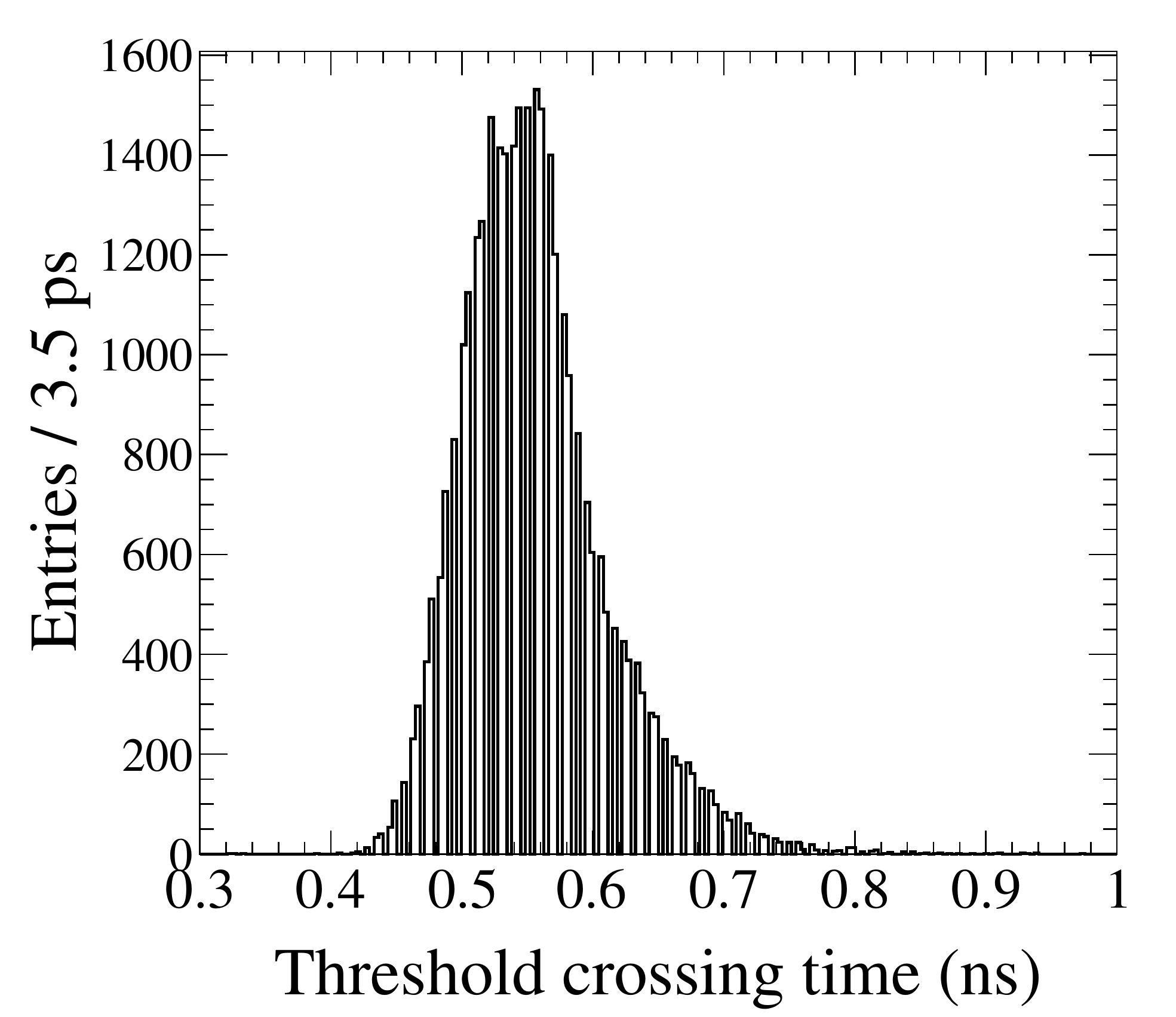}
\hspace{0.0\linewidth}
}
\subfigure[]
{
\label{fig:tdc2Q} 
\includegraphics[width=0.304\linewidth]{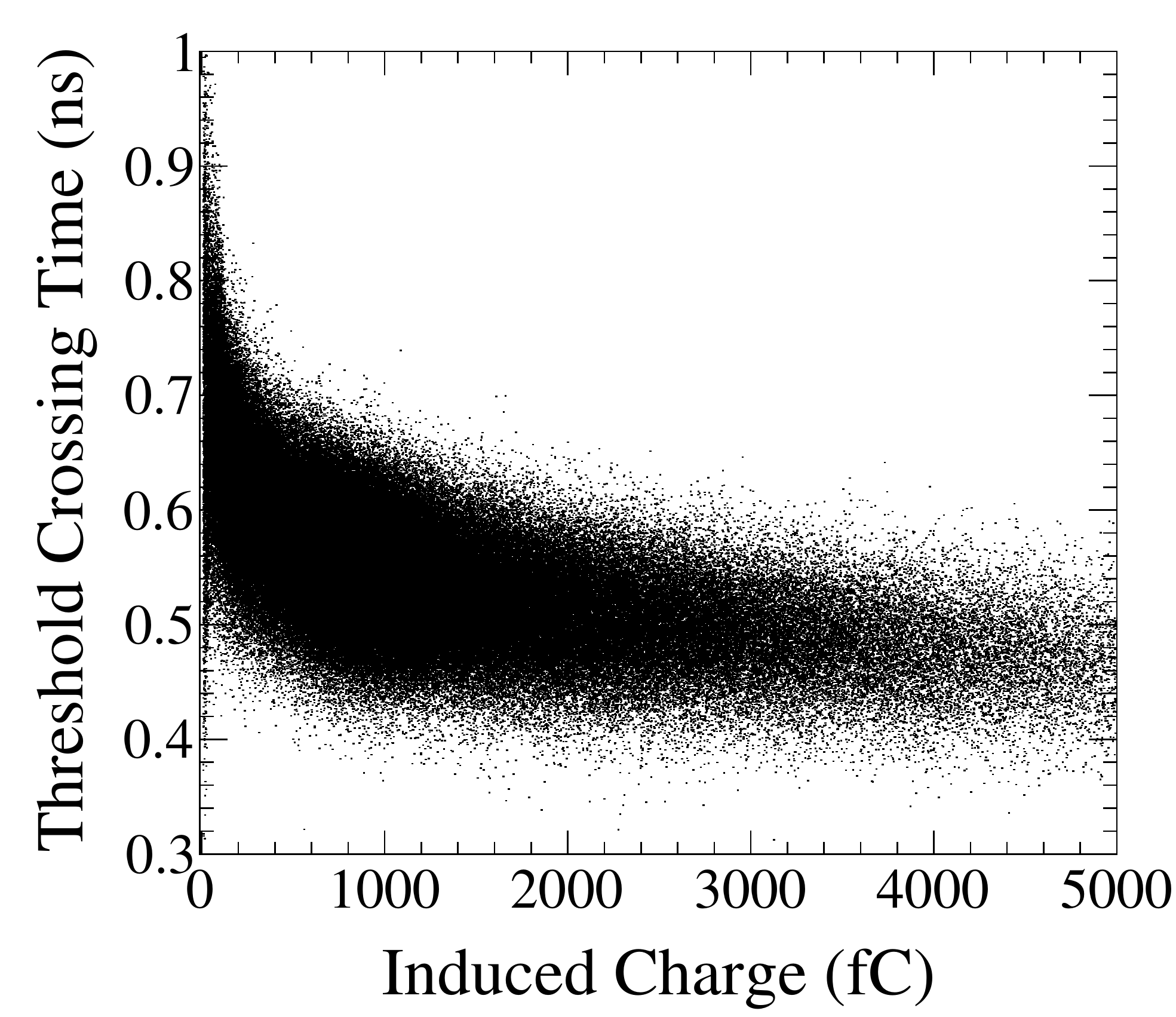}
}
\subfigure[]
{
\label{fig:pulseL}
\includegraphics[width=0.304\linewidth]{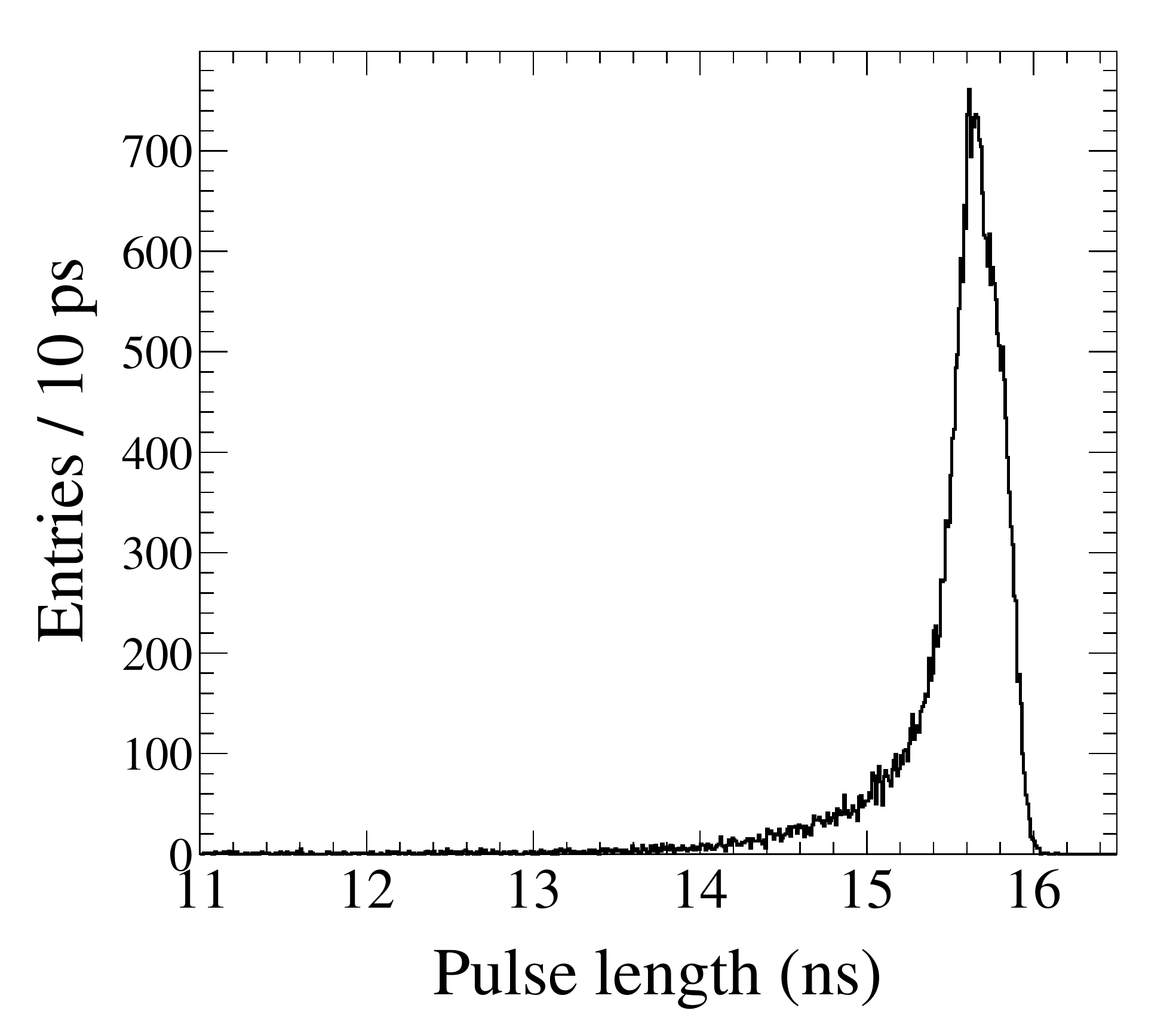}
}
\caption{\label{fig:adc}
(a) Distribution of $t_{\rm thres}$;
(b) charge to time correlation;
(c) distribution of TOT.
}
\end{figure}

In simulation, $t_{\rm flight}$ is defined as the time of the first hit
in the sensitive detector, which is provided by GEANT4.
$t_{\rm thres}$ is determined as the time when the charge 
induced on the strips exceeds the threshold charge.
$t_{\rm prop}$ is calculated using the extrapolated position information
from the  MDC tracks and a propagation speed assumed to be 0.8$c$ 
($c$ denotes the velocity of light in a vacuum.)
Uncertainties of TDC are considered by smearing the measured time 
 with a Gaussian function with the corresponding standard deviation.
 In this study uncertainties from the following sources are considered:
 the jitter of the leading edge of the NINO chips, which depends on the 
 input charge with the relationship empirically obtained; 
 TDC resolution (27 ps); ambiguity of the 12 gaps where the signal is 
 generated (10 ps); electronic components such as cables (20 ps);
 and an additional component with the noise, etc (15 ps).
 The distribution of the resulting $t_{\rm thres}$ is shown in Figure~\ref{fig:thres},
 and the dependence of $t_{\rm thres}$ on the input  charge is 
 illustrated in Figure~\ref{fig:tdc2Q}. 

TOT is calculated from the input charge based on a
conversion function relating the TOT measurement and the 
input charge, which is experimentally obtained.
Then the central value will be smeared by a Gaussian to take uncertainties into 
account: the jitter of the pulse length depending on the input charge,
and an additional uncertainty of 27 ps for TDC resolution. 
Both of them are obtained from experiments.
The TOT distribution  is shown in Figure~\ref{fig:pulseL}.

\section{Comparison with experimental results}
\label{sec:exp}

The data used in this study are $e^+e^-\to e^+e^-$ events
collected by the two prototype modules under
seven high voltages (HV) 
(6700, 6850, 7000, 7150, 7300, 7450 V)  with Threshold of 200 mV,
and four Thresholds  (110, 150, 200, 250 mV) 
with HV of 7000 V.
We generate  MC samples at different working conditions
for comparison with the experimental data.
Both the experimental and MC samples are fully reconstructed
in the same way in BOSS.
In the reconstruction process,
we define $\Delta t$ as the time difference
 between the corrected measured time $t_{\rm meas}$
 and the expected flight  time $t_{\rm exp}$:
 
\noindent
\begin{equation}
\label{eq:deltat}
\Delta t = t_{\rm meas}-t_{\rm exp}.
\end{equation}

\begin{figure}[htp]
\centering
\includegraphics[width=0.38\linewidth]{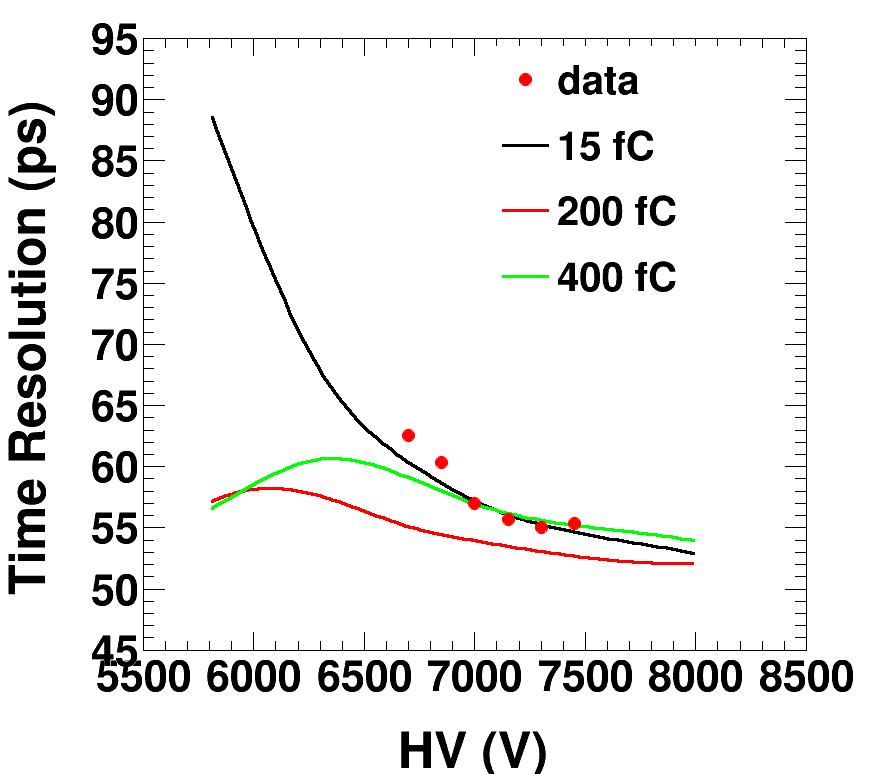}
\qquad
\includegraphics[width=0.38\linewidth]{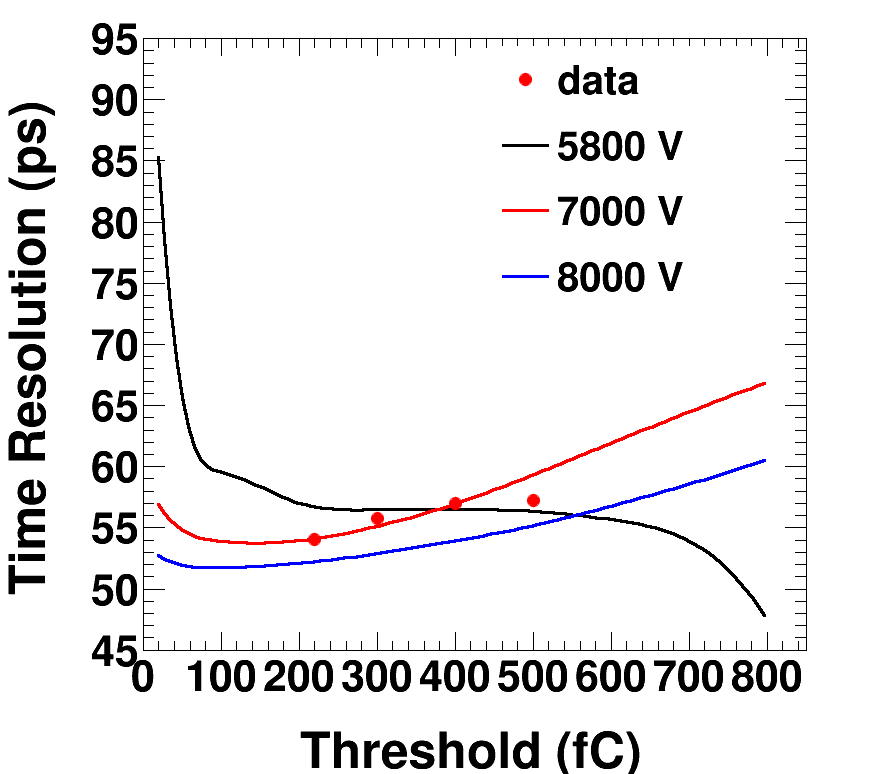}
\caption{\label{fig:res} 
Time resolution versus HV (left) and Threshold (right).
The lines correspond to the simulation results,
while the dots represent the experimental results.}
\end{figure}

Time resolution is obtained by fitting the $\Delta t$ distribution with a Gaussian function
and taking the fitted standard deviation. 
Figure~\ref{fig:res} illustrates the variation of time resolution with HV
in the left plot and with Threshold in the right plot.
The resolution at the favored working condition (HV=7000 V, Threshold=200 mV)
is around 57 ps.
The simulation results are basically consistent with those from the experimental data.
But in the case with a combination of HV<6500 V
and Threshold>400 fC,
the reason of the inverse trend of time resolution variation is still not clear.

\begin{figure}[htp]
\centering
\includegraphics[width=0.38\linewidth]{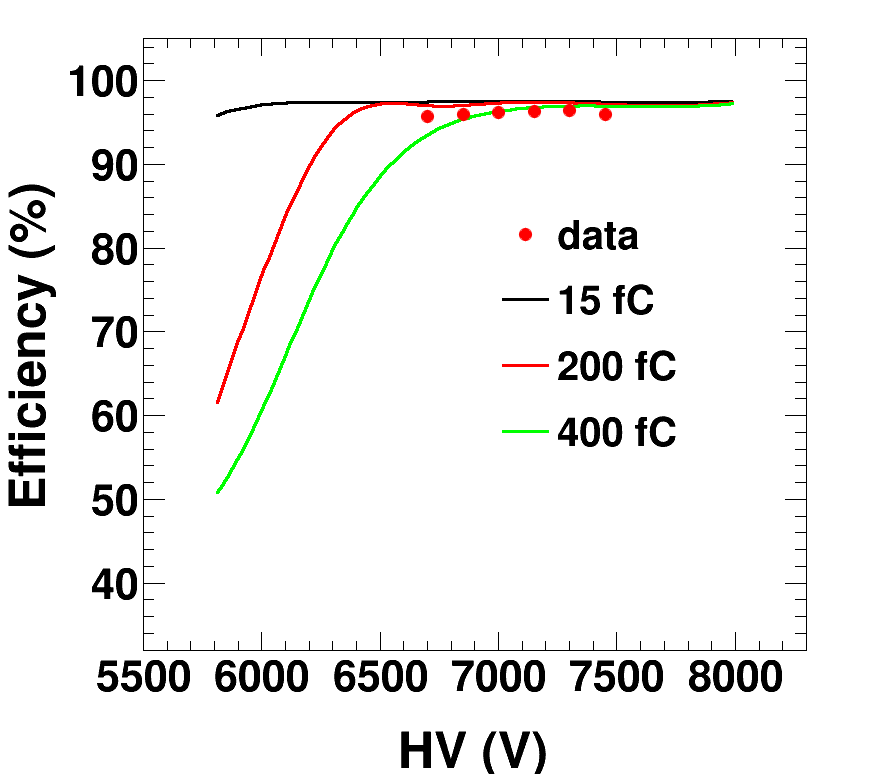}
\qquad
\includegraphics[width=0.38\linewidth]{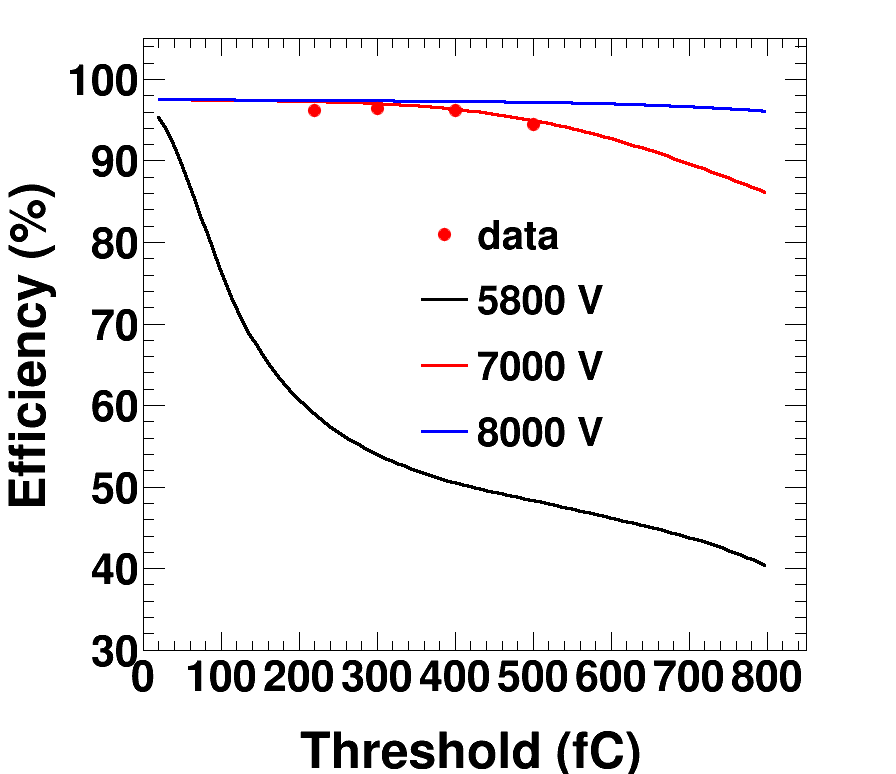}
\caption{\label{fig:eff} 
Detection efficiency versus HV (left) and Threshold (right).
The lines correspond to the simulation results,
while the dots represent the experimental results.}
\end{figure}

Detection efficiency is defined as 

\noindent
\begin{equation}
\label{eq:eff}
\varepsilon = \frac{N_{\rm tof}}{N_{\rm ext}},
\end{equation}

\noindent
where $N_{\rm tof}$ denotes the number of well reconstructed tracks with 
$|\Delta t|<0.8$ ns, and $N_{\rm ext}$ denotes the number of extrapolated MDC tracks.
The efficiency variation versus HV and Threshold is illustrated in Figure~\ref{fig:eff}.
The efficiency plateau is around 97\%.
The simulation results  can well reproduce the behavior 
of the experimental data.
But there is something interesting in the right plot when HV equals 5800 V:
the efficiency decreases sharply when Threshold is small.
This phenomenon indicates that 
under low electric field, the majority of the induced signals are as small
as the 100 fC level.

\section{Conclusion}
\label{sec:con}
A simulation package of MRPC based on Geant4 is developed 
for the upgrade of the BESIII ETOF system. 
Two prototype MRPC modules have been installed into the BESIII 
detector and have taken data during physics runs under a series of 
high voltages and thresholds.
Simulation results of time resolution and detection efficiency under different
working conditions are presented, which agree well with those from the experimental
data. In order to achieve the high precision physics goal of BESIII,
further  study of digitization and MC tunning will be  performed to optimize
the parameters utilized in  simulation based on the upcoming 
experimental data taken by the complete MRPC-based ETOF.




\acknowledgments
This work is supported by National Natural Science Foundation of China (NSFC) 
under Contracts No. 11575225
  and National Key Basic Research Program of China under  Contracts No. 2015CB856706.


\end{document}